\documentclass[10pt,conference]{IEEEtran}
\IEEEoverridecommandlockouts
\usepackage{amsmath,amssymb,amsfonts}
\usepackage{algorithmic}
\usepackage{graphicx}
\usepackage{textcomp}
\usepackage{xcolor}
\usepackage{mathptmx}
\usepackage{siunitx}
\usepackage{float}
\usepackage{epstopdf}
\usepackage{graphicx}
\usepackage{caption}
\usepackage{subcaption}
\usepackage{lipsum}
\usepackage{tikz,cite}
\usepackage{pgfplots}
\usetikzlibrary{decorations.pathreplacing,positioning}
\usepackage{caption,multirow,booktabs}
\usepackage{subcaption}
\usepackage[multiple]{footmisc}

\begin{document}

\title{Reducing safe UAV separation distances with U2U communication and new Remote ID formats 
}

\author{\IEEEauthorblockN{Evgenii Vinogradov$^{1,2}$, 
Sofie Pollin$^2$}
\IEEEauthorblockA{$^1$\textit{Technology Innovation Institute, UAE}; $^2$\textit{Department of Electrical Engineering, KU Leuven, Belgium} \\
Email: evgenii.vinogradov@tii.ae, sofie.pollin@kuleuven.be}
}

\maketitle

\begin{abstract}
As the number of Unmanned Aerial Vehicles (UAVs) in the airspace grows, ensuring that the aircraft do not collide becomes vital for further technology development. In this work, we propose a new UAV Near Mid-Air collision (uNMAC) safety volume taking into account i) Airframe size, ii) Localization precision, iii) UAV speed/velocity, and iv) wireless technology capabilities. Based on uNMAC, we demonstrate that inter-UAV separation distances can be reduced by using UAV-to-UAV (U2U) communication while the safety levels remain unchanged. Moreover, this work shows that next-generation Remote ID messages should contain additional information (i.e., estimated localization error and, for some applications, movement direction). As frequent broadcasting of Remote ID can further reduce the separation distances, we identified 5G NR Sidelink, Wi-Fi, and Bluetooth as suitable candidates for U2U communication.
\end{abstract}


\section{Introduction}
The number of UAV applications is growing. 
The authors of \cite{urban_density18} calculated that by 2035 about 180.000 drone flights per hour would be necessary to shift 70\% of all parcel deliveries in the metropolitan area of Paris into the air. 
On average, it is 63 drones per km$^2$. 
However, there will be busy areas with much higher density.
Such high density causes safety concerns.


Definition of separation distances ensuring safe exploitation of UAVs is of core importance for aerial Conflict Management (CM). Currently, the calculation of these distances is commonly based on methodologies developed for the manned aviation predecessors \cite{Assure,Weinert16,Cook15,Weinert18,Bubbles21}. On the one hand, using the vast experience accumulated during the past century can save much effort on trials and errors. However, on the other hand, now we have a perfect opportunity to critically revise some of the assumptions widely used in civil aviation and adapt them to the new (potentially autonomous or highly automated) agent in the airspace (namely, UAVs).  

Remote ID \cite{Drone_ID} enables transparent registration, flight permission issuing (e.g., by Unmanned Aerial System Traffic management systems - UTM \cite{Vinogradov20, BAURANOV2021}), and, as it contains the aircraft coordinates, ensuring that UAVs are far enough from each other (this is often called “separation provision”).

In this work, we assess the hypothesis that the separation distances may be reduced if UAVs exchange an augmented version of Remote ID containing more information about the aircraft: their size, mobility, and on-board localization/navigation equipment performance. In this paper, we
\begin{itemize}
    \item Overview the State of the Art (SoTA) approaches for defining the separation distances;
    \item Propose a new uNMAC volume taking into account i) Airframe size, ii) Localization precision, iii) UAV speed/velocity, iv) wireless technology capabilities (i.e., broadcasting rate);    
    \item Analyze the contributions of the aforementioned components to the final uNMAC volume;
    \item Propose and compare (in terms of collisions/conflicts per flight hour) three candidates for new Remote ID formats.
    \item Compare several wireless technologies for  U2U Remote ID exchange.
\end{itemize}

\section{Related works}
This section provides an overview of necessary terminology that may be unfamiliar to telecommunication experts. Next, we survey different approaches to defining separation distances for UAVs and describe Remote ID.

\subsection{Used Terminology}
Mid-Air Collision (MAC) happens if two aircraft physically collide. EUROCONTROL defines \cite{eurocontrol} Near Mid-Air Collision (NMAC) as an encounter in which the horizontal separation $d_H$ between two aircraft is less than 150 m (500 ft), and the vertical separation $d_V$  is less than 30 m (100 ft). 
These dimensions are used to calculate other (larger) key volumes and distances used in aviation (e.g., Remaining Well Clear – RWC\cite{Vinogradov20} or distances for Detect-And-Avoid – DAA systems). 

Though NMAC is empirically proven to be suitable for manned aviation, 150x30 m volume is too large to adequately describe a very hazardous situation when two UAVs with a wingspan of 1 m (or less) are involved in the conflict. Unfortunately, we can observe that NMAC is still used for small UAVs \cite{Assure,Weinert16,Cook15,Weinert18}. 
This will lead to overpessimistic estimations of airspace capacity and economic potential of UAV use-cases.


\subsection{State-of-the-Art Overview: Separation Distances}
Since defining various separation distances (e.g., based on NMAC, MAC, and RWC volumes) is critical for multiple UAV exploitation aspects such as demand and capacity balancing, or the design of supporting wireless technologies, this issue has attracted significant attention of many actors (e.g., Federal Aviation Administration - FAA \cite{Assure,Cook15}, NASA \cite{Cook15}, laboratories of national security agencies \cite{Weinert16,Weinert18,Weinert22}, SESAR \cite{Bubbles21}). 
The correspondent contributions are summarized in Table~\ref{Tab:sota}. 

\begin{table}[h]
\centering
\vspace{2mm}
\caption{NMAC STATE OF THE ART OVERVIEW}
\label{Tab:sota}
\begin{tabular}{c||c|c|c|c}
\midrule
\multirow{ 2}{*}{Source}& Reference & Applica-& Communi-&GNSS \\ 
& volume& bility& cation&support\\ \midrule

\multirow{ 2}{*}{ASSURE \cite{Assure}}& NMAC&\multirow{ 2}{*}{U2M}&\multirow{ 2}{*}{NA}&\multirow{ 2}{*}{NA}\\
&(150x30~m)&&&\\
\midrule
SARP \cite{Weinert16,Cook15}& NMAC&U2M&NA&NA\\\midrule
MIT LL\cite{Weinert18}& NMAC &U2M&NA&NA\\ \midrule
MIT LL\cite{Weinert22}& sNMAC &U2U&NA&NA\\ \midrule
\multirow{ 2}{*}{BUBBLES \cite{Bubbles21}}& \multirow{ 2}{*}{MAC} &U2M&via&Upper\\
&&U2U&ground&bound\\ \midrule
\multirow{ 2}{*}{This work}& defined&\multirow{ 2}{*}{U2U}&\multirow{ 2}{*}{U2U}&\multirow{ 2}{*}{Actual}\\
& pairwise&&&\\
\midrule

\end{tabular}
\vspace{-5mm}
\end{table}

The works \cite{Assure,Weinert16,Cook15,Weinert18} are focused on deriving RWC volumes based on NMAC and applicable only for UAV-to-manned (U2M) aircraft conflicts. It was a major concern during the initial phases of introducing UAVs into the National Airspace (NAS).
U2U conflicts attracted attention with a certain delay: for example, \cite{Weinert22,Bubbles21} were published in 2022 and 2021, respectively. Work \cite{Weinert22} from MIT Lincoln Laboratory suggests using small NMAC (sNMAC) volume as a starting point for further separation distance calculations. Analogously to \cite{eurocontrol} where NMAC dimensions roughly reflected the doubled size of an average manned aircraft, the authors of \cite{Weinert22} suggested defining sNMAC  based on the maximum UAV wingspan (7.5 m) found in a database containing UAV characteristics\footnote{http://roboticsdatabase.auvsi.org/home}.

BUBBLES project proposes an approach assuming the Specific Operations Risk Assessment (SORA) risk model \cite{Jarus19} extended to consider UAS operations. 
BUBBLES separation estimations consider Strategic and Tactical Conflict Management \cite{icao_deconf, Vinogradov20} performed by Air and Unmanned Aerial System Traffic Management (ATM and UTM) systems. 
This implies that UAV operators are able to communicate with the ground infrastructure and change their behavior following recommendations. 
This approach considers the presence of a human in the loop, which i) makes the system prone to human error and ii) significantly slows down the system response time and, consequently, increases the required separation distances (i.e., reduces the airspace capacity). 
Another important feature of \cite{Bubbles21} is the consideration of several errors which can be expected in real-life operations. 
The most important one (introducing a 40 m error out of a total of 41 m) is coordinate uncertainty due to errors induced by Global Navigation Satellite Systems (GNSS). 
Note that the value of GNSS error is based on the worst-case performance deduced from the literature.

\textbf{State-of-the-Art Limitations:} U2M separation modeling is well investigated. 
U2U separation is still being defined. 
The solutions available now are either based on non-cooperative UAVs \cite{Weinert22} or consider communication with ground infrastructure \cite{Bubbles21}. 
Moreover, both solutions use several conservative assumptions as described above. 
BUBBLES project offers an interesting solution requiring ground infrastructure (which makes it incompatible with Remote ID). 

Regarding the distances, sNMAC volume \cite{Weinert22} is calculated based \textit{only on a sum of maximum wingspans} (i.e., around 15 m). 
However, it is known that location uncertainty plays an important role in determining the separation \cite{hu2020}. The uncertainty can be caused by a range of factors such as GNSS errors, UAV displacement due to movement, and delays in reporting the location coordinates (i.e., errors induced by wireless communication performance). If all these factors are taken into account, separation distances become mostly dependent on the performance of onboard sensors and communication modules. Note that when the UAVs exchange only their coordinates (as in Remote ID), we must assume upper bounds for all these errors (airframe sizes and localization errors) because the drone can be anywhere (with a certain probability) within the area defined by these errors.

\textbf{Beyond SoTA:} we define uNMAC dimensions considering that relevant information is communicated.
Consequently, we suggest defining the \textit{smallest possible pairwise} uNMAC as a sum of \textit{individual} wingspans of the UAVs. Violation of this volume results in MAC. Final uNMAC is a combination of i) Airframe sizes, ii) reported localization errors, and iii) distance traveled by drones between two coordinate updates.

This work proposes a solution applicable to systems in which safe UAV operations are ensured by separation distances autonomously calculated via onboard processing. Such a solution will be required, for instance, by U3 phase of U-Space where UAVs are expected to benefit from assistance for conflict detection and automated detect and avoid functionalities.

\subsection{State-of-the-Art Overview: Remote ID}
Remote ID is currently not mandatory, however, the final ruling by the FAA and European Union’s Aviation Safety Agency (EASA) on Remote ID will require most drones operating within the US/EU airspace to have Remote ID installed. All drones will be required to have Remote ID capabilities by September 2023 and January 2024 to have access to the national airspace of the US and all the EU member states\footnote{A number of exceptions will exist: for example in the US, Remote ID will not be mandatory for Visual Line of Sight operations performed by educational institutions in certain areas. In the EU, it will not be required for drones weighing less than 250 grams (including payload) and having no camera or other sensor that can capture personal data.}.

The key requirements to Remote ID can be listed as \cite{Drone_ID}:
\begin{itemize}
    \item Remote ID messages should be broadcasted directly from the UAV via radio frequency broadcast (likely via Wi-Fi or Bluetooth technology\footnote{Initially, the FAA considered a possibility of using ground infrastructure and ADS-B but, due to a number of problems well summarized in \cite{Vinogradov20}, these technologies were excluded from the final version of the recommendations.}), and be compatible with existing personal wireless devices.
    \item Remote ID message includes i) UAV ID (serial number of UAV or the session ID); ii) latitude/longitude, altitude, and velocity of UAV; iii) latitude/longitude and altitude of Control Station; iv) emergency status, and v) time mark.
    \item Range of the remote ID broadcast may vary, as each UAV must be designed to maximize the range at which the broadcast can be received.
    \item Remote ID broadcast cannot be disabled by the operator. Moreover, UAVs must self-test so the drone cannot take off if the Remote ID is not functioning.
\end{itemize}
\textbf{Beyond SoTA:}
We suggest including other relevant fields in addition to the information in the standard Remote ID message. In this work, we evaluate candidates:
\begin{itemize}
    \item \textbf{Candidate~1:} i) upper bound for airframe size; actual ii) localization error and iii) speed.
    \item \textbf{Candidate~2:} actual i) airframe size, ii) localization error, iii) speed.
\item \textbf{Candidate~3:} actual i) airframe size, ii) localization error, iii) speed and direction of movements (following the claims of \cite{hu2020} where the direction was used).
\end{itemize}
The candidates are compared with standard Remote ID messages and MAC (e.g., when there is no errors/communication delays/location uncertainty).

\section{System Model}\label{sec:system}
\begin{figure}
\centering
       \includegraphics[width=0.6\columnwidth   ]{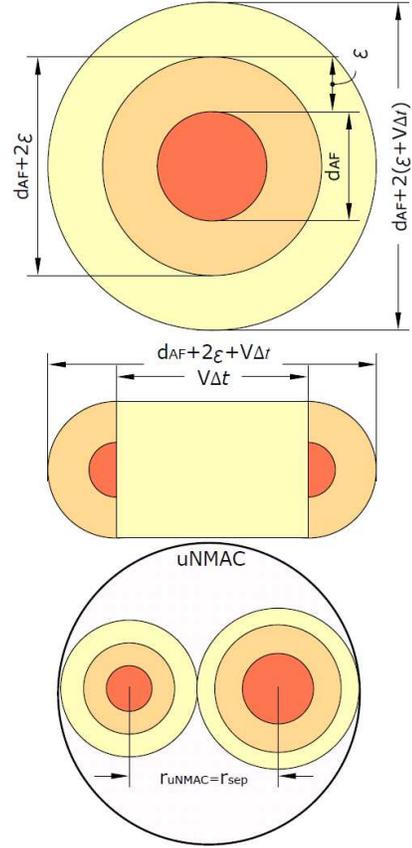}
       \centering
		\caption{Top: UAV location uncertainty area consists of i) Airframe, ii) Localization error, iii) Displacement (in an unknown direction) between two location reports. Middle: UAV uncertainty area (the mobility direction is known). Bottom: UAV Near Mid-Air Collision (uNMAC) and the correspondent (minimal) separation distance. }
		\label{Fig:Idea}
  \vspace{-4mm}
\end{figure}

Consider a UAV of airframe size $d_{AF}$ moving with speeds $V$ in a certain direction. The aircraft is equipped with i) Self-localization (e.g., GPS) and ii) wireless communication modules. The localization module reports the UAV coordinates with an error of $\pm \epsilon_{}$. We assume that the airframe and localization errors are symmetric around the center of the drone. We assume that location update and communication (broadcasting) rates $\Delta t_{LOC}=\Delta t_{COM}$ are the same and denoted as $\Delta t$. In other words, once the location (coordinates) is updated, this update is immediately broadcasted. The aircraft travels for a distance $V\Delta t$ between two location updates. If there is no information about the movement direction, the UAV can be anywhere in the area shown in Fig.~\ref{Fig:Idea}, top. If the direction is known, the shape of the uncertainty area changes as shown in Fig.~\ref{Fig:Idea}, middle. 

When there are two (or more) drones in the airspace, safe UAV operation is possible only if the separation distance $r_{sep}$ ensures that the uncertainty areas around the drones do not overlap (Fig.~\ref{Fig:Idea}, bottom). 
In this work, we assume a single altitude slice of the airspace and, consequently, we focus on the horizontal separation. 
This 2D study will be extended to a 3D case in the following works. No collision avoidance technique is implemented (to assess only the Remote ID influence).

\underline{Airframe size:}
The authors of \cite{Weinert22} used a database containing the characteristics of UAVs. Following the authors, we model the airframe size as a uniformly distributed random variable not exceeding $d_{AF}^{max}=7.5$~meters.

\underline{Localization error:}
Though many possible solutions exist for UAVs' self-localization (e.g., based on visual Simultaneous Localization and Mapping - SLAM \cite{Vinogradov21}), this work is based on using GPS as it is the most common solution used nowadays. The conclusions of this paper can be easily extended to SLAM or other GNSS such as Galileo, GLONASS, and BeiDou by using the range estimation errors reported by these systems. 

Conventionally, the positioning error in GPS is modeled as a normally distributed random variable $\mathcal{N}(0,\sigma^2)$ with zero mean and the standard deviation $\sigma$ depending on several factors. As stated by the GPS performance standards \cite{GPS_SPS_2020}, the positioning error can significantly differ depending on the age of data (AOD), which is the elapsed time since the Control Segment generated the satellite clock/ephemeris prediction used to create the navigation message data upload. Different errors listed in the document (see \cite{GPS_SPS_2020}, Table 3.4-1) are summarized in Table~\ref{tab:gps}. Note that in the table, we give $3\sigma$ to cover 99.7\% of possible errors. In some cases \cite{Bubbles21}, the upper bound of the GPS positioning error is set to 40 meters.
\begin{table}[]
     \caption{GPS ACCURACY STANDARDS \cite{GPS_SPS_2020}}   \centering
    \begin{tabular}{c|c}
        $3\sigma$, m & Accuracy standard \\
        \hline
         5.7& Normal Operations at Zero AOD\\
         10.5&Normal Operations over all AODs\\
         13.85&Normal Operations at Any AOD\\
         30&Worst case, during Normal Operations\\
    \end{tabular}
\vspace{-6mm}
    \label{tab:gps}
\end{table}

\underline{UAV velocity:} 
An aircraft has many different defined airspeeds; these
are known as V-speeds. In \cite{Weinert18}, two V-speeds were leveraged when defining the representative UAV: cruise $V_C$, the speed at which the
aircraft is designed for optimum performance, and the maximum operating speed $V_{max}$. Under an assumption that most UAV operators will leverage the vendor-recommended performance guidelines, the authors suggested modeling the UAV airspeeds by means of a Gaussian distribution $\mathcal{N}(\mu_v,\sigma^2_v)$ with $\mu_v=V_C$ and a heuristically defined standard deviation:
\begin{equation}\label{eq:speed_sigma}
    \sigma_v = \frac{V_{max}-V_c}{3}.
\end{equation}

Based on the aforementioned database, the authors defined four UAV categories based on the correspondent speeds and maximum gross takeoff weights (MGTOW), see Table~\ref{tab:velocity}. 
\begin{table}[t]
    \centering
    \vspace{1mm}
        \caption{REPRESENTATIVE UAV CATEGORIES}
    \begin{tabular}{c|c|c|c|c}
        &1&2&3&4  \\
        \hline
        MGTOW, kg&0-1.8&0-9&0-9&9-25\\
        Mean cruise speed $V_c$, m/s&12.9&10.3&15.4&30.7\\
        Max airspeed $V_{max}, m/s$&20.6&15.4&30.7&51.5\\
        
    \end{tabular}
    \label{tab:velocity}
\end{table}

\underline{Location update and wireless broadcasting rates:}
GPS module producers offer a wide range of equipment supporting various position update rates $\Delta t$. The update rate can go as high as 50~Hz (e.g., Venus838FLPx by SkyTra is able to provide localization/position updates every $\Delta t$= 20~ms) while many accessible consumer grade modules mostly offer between 0.2-8 Hz update rates.
An overview of wireless communication technologies is provided in Table~\ref{tab:techs}. 
\begin{table}[t]
    \centering
        \caption{WIRELESS TECHNOLOGIES FOR UAV-TO-UAV COMMUNICATIONS}
    \begin{tabular}{c|c|c|c}
    Technology & {Range} & {Update rate}& $\Delta t_{com}$  \\
    \hline
    Bluetooth LE & 50~m & 100~Hz& 10~ms \\
    Bluetooth & 100~m & 100~Hz& 10~ms \\
    LoRa \cite{Vinogradov20} & 10~km & 0.2~Hz& 5 s  \\
    FLARM \cite{8514548}  & 10~km & 0.33~Hz & 3 s  \\
    Wi-Fi SSID \cite{9133405}&1~km&60~Hz& 16~ms \\
    5G NR Sidelink \cite{9345798}&1~km& up to 1-8~kHz& 0.125-1~ms\\
    \end{tabular}
\vspace{-6mm}
    \label{tab:techs}
\end{table}

\section{Proposed uNMAC definition}\label{sec:uNMAC}

Assuming that the movement direction is unknown, we may express the diameter of the circular uncertainty area around UAV $i$ (see Fig.~\ref{Fig:Idea}, top) as
\begin{equation}\label{eq:uNMAC_1}
    d_i = d_{AF,i} + 2(\epsilon_{i} + V_i\cdot\Delta t),
\end{equation}
where 
$2\epsilon_{}$ with the subscript denotes the increase of the uncertainty is caused by the GPS module instrumental error. As Candidate 3 relies on knowing the movement direction, we may rewrite \eqref{eq:uNMAC_1} as 
\begin{equation}\label{eq:uNMAC}
    d_i = d_{AF,i} + 2\epsilon_{i} + \vec{V_i}\cdot \Delta t,
\end{equation}
where $\vec{V_i}$ contains both speed and direction of movement
. The resulting area is shown in the middle part of Fig.~\ref{Fig:Idea}.

As the UAV can be anywhere the area calculated by \eqref{eq:uNMAC_1}, uNMAC should include the areas around UAV$_i$ and UAV$_j$ (see Fig.~\ref{Fig:Idea}, bottom). uNMAC becomes a circle with a diameter
\begin{equation}\label{eq:uNMAC_circle1}
    d_{uNMAC}^{ij} = d_{AF,i}+d_{AF,j}+2(\epsilon_{i}+ \epsilon_{j}+ (V_i+ V_j)\cdot\Delta t).
\end{equation}

For \eqref{eq:uNMAC}, we may rely on the relative movement of the aircraft (it is equal to zero for parallel trajectories):
\begin{equation}\label{eq:uNMAC_circle2}
    d_{uNMAC}^{ij}= d_{AF,i}+d_{AF,j}+2(\epsilon_{i}+ \epsilon_{j})+V_{rel}\cdot\Delta t.
\end{equation}
Based on \eqref{eq:uNMAC_circle1} and \eqref{eq:uNMAC_circle2}, we may calculate the separation distances ensuring that UAV$_i$ and UAV$_j$ do not collide as
\begin{equation}\label{eq:separation}
r_{uNMAC}=\frac{d_{uNMAC}}{2}.
\end{equation}
When two UAVs are closer than $r_{uNMAC}$, there is a non-zero probability that the UAVs physically collide.
Note that a Midair Collision happens if the distance between the centers of the two UAVs is smaller than 
\begin{equation}
    r_{MAC}=\frac{d_{AF,i}+d_{AF,j}}{2}.
\end{equation} 

To ensure safe exploitation of aircraft, inter-UAV separation must satisfy the condition $r_{sep} \geq r_{uNMAC}$. It is obvious that if the errors are significant and $r_{uNMAC}$ becomes much larger than $r_{MAC}$, the airspace will not be used rationally. This may happen, for instance, if we do not have information about the localization precision (as in Remote ID) and must consider the upper bound for this error.  

\underline{Airframe size:}
We may model the distribution of the sum of two half airframe sizes $d_{AF}=(d_{AF,i}+d_{AF,j})/2$ as the triangle distribution with the density function
\begin{equation}\label{eq:AF}
  f_{AF}(x)=\begin{cases}
    \frac{x}{AF_{max}^2/4}, & \text{if $0<x<\frac{AF_{max}}{2}$}\\
    \frac{AF_{max}-x}{AF_{max}^2/4}, & \text{if $\frac{AF_{max}}{2}\leq x<AF_{max}$}\\
    0, & \text{otherwise}.\\
  \end{cases}
\end{equation}
 
\underline{Localization error:}
In this paper, we are not interested in the instrumental error per se; we rather target calculating the uncertainty area expansion due to the localization error and its influence on $r_{uNMAC}$. Any localization error $X$ causes an increase of the uncertainty area $\epsilon_{}=|X|$, where $X$ follows a normal distribution $\mathcal{N}(0,\sigma^2)$ as described in Section~\ref{sec:system}. Consequently, $\epsilon_{}$ follows the Half-Normal distribution:
\begin{equation}
    f(x,\sigma)=\frac{\sqrt{2}}{\sigma\sqrt{\pi}}\exp{-\frac{x^2}{2\sigma^2}},  \quad   x\geq0.
\end{equation}
Based on this distribution, we may derive the probability density function of $\epsilon_i+\epsilon_j$ contributing to \eqref{eq:separation} as
\begin{eqnarray}\label{eq:loc}
    f(x)=\frac{1}{\sqrt{\sigma^2_i+\sigma^2_j}} \sqrt{\frac{2}{\pi}} \cdot \exp{\Big(-\frac{x^2}{2(\sigma^2_i+\sigma^2_j)}}\Big) \times \nonumber\ \\
    \Bigg[ \mathrm{erf} \Bigg(\frac{\sigma_i x}{\sqrt{2}\sigma_j \sqrt{\sigma_i^2+\sigma^2_j)}}\Bigg)+
    \mathrm{erf} \Bigg(\frac{\sigma_j x}{\sqrt{2}\sigma_i \sqrt{\sigma_i^2+\sigma^2_j)}}\Bigg) \Bigg],
\end{eqnarray}
where $\sigma_{i,j}$ are the standard deviations of the errors estimated by UAVs $i$ and $j$ respectively, and $\mathrm{erf}(\cdot)$ is the error function.

\underline{UAV velocity:} When the movement direction is not known, we rely on the maximum relative speed $V_{rel}^{max}$ which can be calculated as a sum of two independent normal variables, consequently, it follows the Gaussian distribution $\mathcal{N}(\mu_{v1}+\mu_{v2},\sigma^2_{v1}+\sigma^2_{v2})$, where $\mu_{v1, v2}$ and $\sigma^2_{v1,v2}$ are the speed distribution parameters for the UAVs involved in the potential conflict.

However, when the direction is reported, 
the relative speed can take values $-V_{rel}^{max}\geq V_{rel}\geq V_{rel}^{max}$. The relative movement direction can be modeled as a uniformly distributed random variable $Y$ taking values from $-\pi$ to $\pi$. Consequently, the direction-dependent velocity can be modeled as
\begin{equation}
    \vec{V}= V_{rel}^{max} \cdot cos(Y),
\end{equation}
where $V_{rel}^{max}$ can be modeled as for the unknown direction case. 

\underline{Location update and wireless broadcasting rates:}
As $\Delta t$ in \eqref{eq:uNMAC_1}-\eqref{eq:separation} is linked to speed, the broadcasting/localization update rates influences the distribution of the mobility-induced uncertainty. As speeds are modeled by a normal random variable, the distribution of $V \Delta t$ is also described by a Gaussian distribution with mean $\mu=\Delta t (\mu_{v1}+\mu_{v2})$ and variance $\Delta t^2 (\sigma_{v1}^2+\sigma^2_{v2})$.

\section{Results}
\subsection{Contributions of the uNMAC components}
\begin{figure}
    \centering
    \includegraphics[trim={0.3cm 0 0 0.48cm },clip, width=1.1\columnwidth]{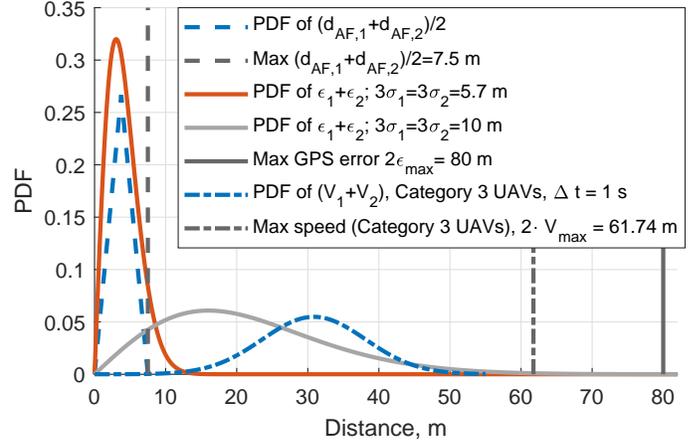}
    \caption{Probability density functions of the $r_{uNMAC}$ components. Reporting the localization error is highly beneficial.}\label{fig:components}
    \vspace{-5mm}
\end{figure}

To assess the sensitivity of $r_{uNMAC}$ distances, the contributions of the three individual components are plotted in Fig.~\ref{fig:components}. It is evident that using actual $d_{AF}, \epsilon_{}$, and $V\Delta t$ reduces the size of the total uNMAC area. Notably, when GPS is used for localization, reporting the actual error is highly beneficial: while the upper bound error is 80~m, we can achieve the mean errors of 3.03~m and 7.37~m for Zero and all AODs ($3\sigma=5.7$~m and $13.75$~m, see~Table~\ref{tab:gps}), respectively. For the same AODs, the localization errors do not exceed 9.4~m and 22.88~m with a probability of 99.9\%.

The improvement brought by using actual airframe sizes is not that significant: while the sum of two largest airframe sizes requires a separation of 7.5~m, the mean value is 3.75~m. Note that we modeled the airframe sizes by a uniform distribution. The benefits of using actual numbers instead of the maximal ones can increase if small-size UAVs are predominant in the airspace. However, it is impossible to achieve the $60-70$~m improvement offered by reporting actual localization accuracy.

\begin{figure}
    \includegraphics[trim={0.6cm 0 0 0.48cm },clip, width=1.1\columnwidth]{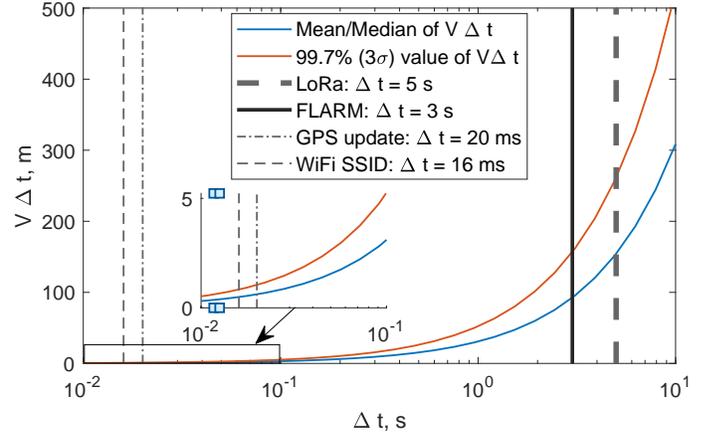}
           \centering
    \caption{Increasing localization/broadcasting rate can significantly reduce the mobility-induced error. Note that 5G NR Sidelink (min $\Delta t$=0.125~ms and Bluetooth (min $\Delta t$=10~ms) can offer sub-1~meter error.}\label{fig:speed}
    \vspace{-6mm}
    \end{figure}

As it follows from the representation shown in Fig.~\ref{Fig:Idea}, knowing the movement direction can result in $r_{uNMAC}$ ranging  
\begin{equation}
    a \leq r_{uNMAC} \leq a+\Delta t (V_1+V_2),
\end{equation}
where $a=r_{MAC}+\epsilon_{1}+\epsilon_{2}$.
Consequently, for $\Delta t \to 0$, the impact of knowing mobility direction diminishes. Moreover, as stated in Section~\ref{sec:uNMAC} and demonstrated by Fig.~\ref{fig:speed}, we may significantly reduce the mobility-induced uncertainty (for both known and unknown movement directions) by sending the updates more often (i.e., decreasing $\Delta t$). Using 5G NR Sidelink, Bluetooth, or WiFi SSID can be recommended, while FLARM and LoRA result in overly conservative separation distances. 
\subsection{Remote ID candidates evaluation}
\subsubsection{Flight simulator}
In order to assess the influence of the uNMAC volume on the expected airspace capacity, we created a 2D flight simulator considering that all UAVs are flying at the same altitude. Though the practical airspace capacity depends on the number of used flight altitudes, the result obtained for a single height can provide enough information to assess the benefits of sharing more information between the UAVs. 

The simulator considers a squared  area $S$ of 10~km$^2$. Depending on the UAV density $\lambda$, we generate $N=\lambda\cdot S$ UAVs. For each UAV, we follow the steps i) the start and end points of the UAV trajectory are generated from the uniform distribution; ii) the velocities are drawn from $\mathcal{N}(\mu_v,\sigma^2_v)$ using the parameters taken from Table \ref{tab:velocity} (the categories are picked randomly) and \eqref{eq:speed_sigma}, iii) the UAV airframe size is sampled as it was described in Section~\ref{sec:system}.

Next, in a pairwise manner, we calculate the closest points of approach for all trajectories. Trajectory pairs with $r_{sep}\leq d_{AF, max} + 2\cdot(\epsilon_{max} + V_{max} \Delta t)$ are identified. For this check, we use the values for Category 4 UAVs (the fastest ones) and $\Delta t = 1$~s; $d_{AF, max}=7.5$~m, $\epsilon_{max}=40$~m. These trajectories are checked in for collisions/conflicts (i.e., the situations when the time-dependent horizontal miss distance between two UAVs is smaller than MAC/uNMAC). Note that all the Remote ID candidates described above are used to calculate uNMAC volumes defined by their configurations (see Table~\ref{tab:sim}). 
Finally, we estimate collisions (MAC) and conflicts (uNMAC violations creating hazardous situations) per flight hour $C(\lambda)$. 

\begin{table}[h]
    \centering
        \caption{SIMULATION PARAMETERS}
    \begin{tabular}{c|c|c|c|c}
        &Remote ID &Candidate 1&Candidate 2&Candidate 3\\
        \hline
        AF& max: 7.5~m&max: 7.5~m&actual&actual\\
        GPS& max: 80~m& $3\sigma=5.7$~m&$3\sigma=5.7$~m&$3\sigma=5.7$~m\\
        Mobility& speed &speed&speed&+direction\\
        \hline
       runs&\multicolumn{4}{c}{10.000.000 trajectories (520 days of uninterrupted flight)}
    \end{tabular}
    \label{tab:sim}
\end{table}

\subsubsection{Candidate comparison}
\begin{figure}[t!]
    \centering
\begin{subfigure}{.455\textwidth}
    \includegraphics[trim={0.4cm 0 0 0cm},clip, width=1.1\columnwidth]{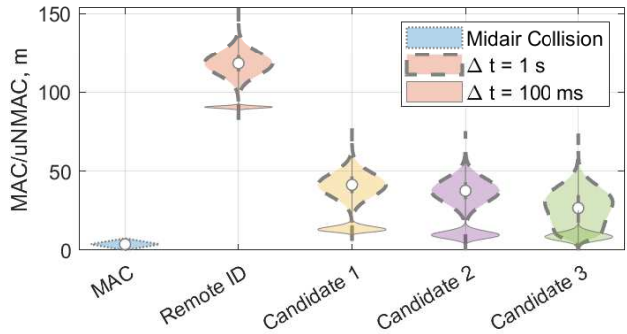}
    \caption{uNMAC distances can be reduced by 1) broadcasting the localization accuracy and 2) decreasing the time between the updates.}\label{fig:uNMAC}
\end{subfigure}
\\
\begin{subfigure}{.455\textwidth}
    \includegraphics[width=1\columnwidth]{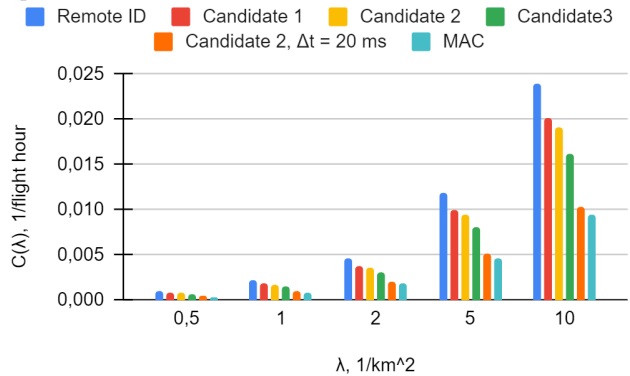}
           \centering
    \caption{Conflicts/collisions per flight hour.}\label{fig:fph}
    \end{subfigure}
    \caption{Comparison of the Remote ID candidates.}\label{fig:candidates}
    \vspace{-6mm}
\end{figure}

Compared to Remote ID, all the suggested candidates offer an opportunity to reduce uNMAC and, consequently, separation distances, as demonstrated by the violin plot shown in Fig.\ref{fig:uNMAC}. Mostly, this is achieved by broadcasting the localization error reported by the dedicated module (GPS in our case). Further improvement is possible if the broadcasting rate is increased. In the figure, we show that uNMAC size distributions become $\sim10$ times more narrow, and their mean is significantly decreased. Fig.\ref{fig:uNMAC} also confirms that Candidate 3 can overperform the other candidates (i.e., knowing mobility direction is useful) only for large $\Delta t$. When $\Delta t = 100$~ms, Candidates 2 and 3 perform equally.   

The UAV density-dependent collision/conflict probability behavior is shown in Fig.~\ref{fig:fph}. Using Remote ID results in detecting $\sim2.5$ times more conflicts than actual collisions (i.e., MAC). This can create situations where DAA systems will be used when it is not necessary. When the resolving capacity of such systems is achieved, the only solution is to lower the UAV traffic density. Again, Candidate 3 is the best when $\Delta t =1$~s; however, when $\Delta t=20$~ms, using Candidate 2 results in a minimal increase of the detected conflicts if compared with MAC. Note that Candidate 2 is our preferred option for small $\Delta t$ since it contains less information than Candidate 3 while offering the same performance.
\section{Conclusions}
In this paper, we proposed a new uNMAC definition taking into account i) Airframe size, ii) Localization precision, iii) UAV speed/velocity, iv) wireless technology capabilities (i.e., broadcasting rate). Based on this definition, analyzed the possible evolution of Remote ID, allowing for reducing the UAV separation distances while preserving the same level of safety. Our main findings can be summarized as follows:
\begin{itemize}
    \item Next-generation Remote ID should additionally contain localization errors;
    \item Information about movement direction is relevant only for relatively low broadcasting rates (1 Hz and lower);
    \item Increasing the broadcasting rate is highly beneficial though it should be aligned with the localization update rate. Consequently, we recommend using Bluetooth (for short flights), Wi-Fi (e.g., by embedding Remote ID in SSID as in \cite{9133405}) or 5G NR Sidelink.

\end{itemize}
\bibliographystyle{IEEEtran}
\bibliography{main}

\vspace{12pt}
\color{red}

\end{document}